\documentclass[conference]{IEEEtran}
\IEEEoverridecommandlockouts
\usepackage{cite}
\usepackage{amsmath,amssymb,amsfonts}
\usepackage{algorithmic}
\usepackage{graphicx}
\usepackage{textcomp}
\usepackage{xcolor}
\usepackage{fancyhdr}

\usepackage[ruled,vlined]{algorithm2e}
\usepackage{booktabs}
\usepackage{enumitem}
\usepackage[italicdiff]{physics}
\usepackage{amsthm}
\theoremstyle{definition}
\newtheorem{definition}{Definition}

\def\BibTeX{{\rm B\kern-.05em{\sc i\kern-.025em b}\kern-.08em
    T\kern-.1667em\lower.7ex\hbox{E}\kern-.125emX}}

\begin{document}

\title{Link Prediction for Temporally Consistent Networks}


\author{\IEEEauthorblockN{Mohamoud Ali}
\IEEEauthorblockA{\textit{School of Computing and Engineering} \\
\textit{University of Missouri-Kansas City}\\
Kansas City, United States \\
mali@mail.umkc.edu}
\and
\IEEEauthorblockN{Yugyung Lee}
\IEEEauthorblockA{\textit{School of Computing and Engineering} \\
\textit{University of Missouri-Kansas City}\\
Kansas City, United States \\
leeyu@umkc.edu}
\and
\IEEEauthorblockN{Praveen Rao}
\IEEEauthorblockA{\textit{College of Engineering} \\
\textit{University of Missouri-Columbia}\\
Columbia, United States \\
praveen.rao@missouri.edu}

}

\maketitle
\thispagestyle{fancy}

\begin{abstract}
Dynamic networks have intrinsic structural, computational, and multidisciplinary advantages. Link prediction estimates the next relationship in dynamic networks. However, in the current link prediction approaches, only bipartite or non-bipartite but homogeneous networks are considered. The use of adjacency matrix to represent dynamically evolving networks limits the ability to analytically learn from heterogeneous, sparse, or forming networks. In the case of a heterogeneous network, modeling all network states using a binary-valued matrix can be difficult. On the other hand, sparse or currently forming networks have many missing edges, which are represented as zeros, thus introducing class imbalance or noise. We propose a time-parameterized matrix (TP-matrix) and empirically demonstrate its effectiveness in non-bipartite, heterogeneous networks. In addition, we propose a predictive influence index as a measure of a node's boosting or diminishing predictive influence using backward and forward-looking maximization over the temporal space of the n-degree neighborhood. We further propose a new method of canonically representing heterogeneous time-evolving activities as a temporally parameterized network model (TPNM). The new method robustly enables activities to be represented as a form of a network, thus potentially inspiring new link prediction applications, including intelligent business process management systems and context-aware workflow engines. We evaluated our model on four datasets of different network systems. We present results that show the proposed model is more effective in capturing and retaining temporal relationships in dynamically evolving networks. We also show that our model performed better than state-of-the-art link prediction benchmark results for networks that are sensitive to temporal evolution.
\end{abstract}

\begin{IEEEkeywords}
Dynamic Networks, Link Prediction, Graph Modeling, Neural Networks
\end{IEEEkeywords}

\section{Introduction}
Link prediction estimates the next relationship in dynamic networks. This research area continues to attract interest~\cite{martinez2017survey} for its intrinsic applicability to various problem domains, including language translation using word sequence networks~\cite{wijaya2017learning}, image classification using pixel networks~\cite{lu2017simultaneous}, and author relationship using citation networks~\cite{wang2016structural}. Its multidisciplinary applications are evident in a range of AI fields, including molecular biology, genomics, social networks, signal processing, and acoustics~\cite{zhao2017link}. However, dynamic networks present significant challenges with varying degrees of learning complexity due to their different structural properties. Continuously changing contexts in local networks are one of the learning complexities~\cite{albrecht2018autonomous}. Learning the structure-function relationship of dynamic networks is complex. It entails learning models that are adaptive to the changing network topology. In addition, the evolution of time, neighboring networks, and other external factors may influence network behavior. 

Recent work tackled network modeling with temporal aspect and/or with global network features. A link prediction approach that considers global network features was introduced in~\cite{gao2011temporal}. Later, a network propagation model that captured temporal features and weights of the evolutionary network over multiple timesteps was introduced~\cite{yu2017link}. However, the spatial and temporal consistency presented in~\cite{yu2017link} only works on the assumption that the nodes within a network are similar. Therefore, it does not work with structurally heterogeneous networks whose nodes have dissimilar features. A heterogeneous network or graph $G = (V, E)$ with type(s) of vertices $T_v$ and type(s) of edges $T_e$ is defined as a graph in which $|T_v| > 1$ or $|T_e| > 1$~\cite{cai2018comprehensive}. Conversely, a homogeneous graph is one in which $|T_v| = |T_e| = 1$.

\begin{figure*}
  \includegraphics[scale=0.22]{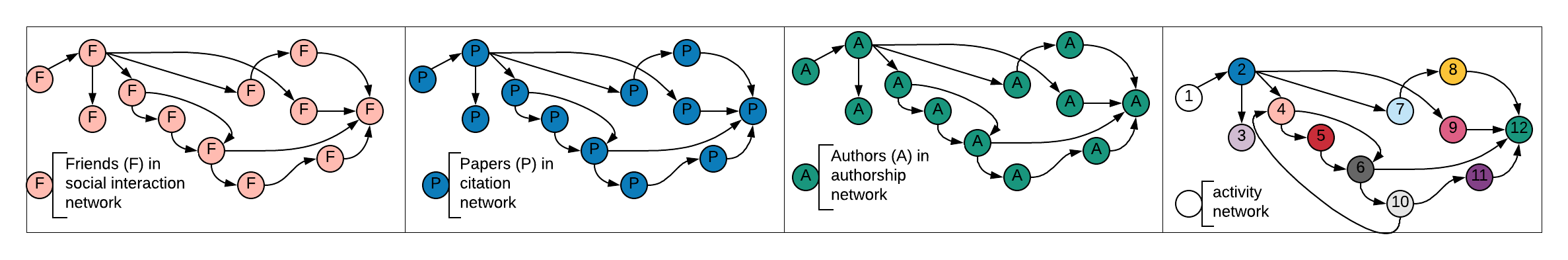}
  \caption{An example of different link prediction network types: Existing link prediction solutions focus only on either bipartite networks or non-bipartite networks with similar node features such as social interaction (F), citation (P), and authorship (A) networks. Colors and node identifiers represent the heterogeneity characteristics of network nodes. Activity networks such as time-evolving business workflows consist of correlated but heterogeneous activities or nodes. Node identifiers for the activity network are shown in Table \ref{tab:activities}.}
  \label{fig:networktypes}
\small
\end{figure*}

To the best of our knowledge, current link prediction solutions only address either bipartite networks or non-bipartite networks with structurally and temporally homogeneous nodes. Figure \ref{fig:networktypes} shows an example of existing link prediction solutions in contrast with heterogeneous activity networks. 

The key contributions of this paper are as follows.

\begin{itemize}
\item
We propose a novel model, called temporally parameterized network model (TPNM), for canonically representing heterogeneous time-evolving activities.  We argue this can inspire further research in practical deep learning applications, such as business workflow optimizations, viral marketing with time-relevant advertisement placement, and talent recruitment process recommendations. 
\item
We introduce a new time-parameterized matrix (\textit{TP-matrix}) as an alternative to adjacency matrix. We empirically show that \textit{TP-matrix} is more suitable for heterogeneous, time-evolving networks.
\item
In support of continuous prediction capability, we design a new approach that distinguishes and separately learns from the time-function relationship and structure-function relationship. This new approach addresses a limitation in current link prediction methods that consider the network evolution as a function of continuous snapshots. 
\end{itemize}

The rest of the content is organized as follows. We review existing link prediction approaches in Section \ref{sec:related} before we broaden our discussion to related AI research areas. Our proposed model and algorithms are presented in Section \ref{sec:methods}. The experiment design, datasets, and results are covered in Section \ref{sec:experiments}. In Section \ref{sec:applications}, we discuss our work's real-world application prospects and how it extends the benefits of using dynamic networks and AI to new domain areas. Section \ref{sec:conclusions} concludes the paper. 

\section{Related Work}\label{sec:related}
Huang et al. \cite{huang2008framework} studied relationship mining for subsequent events. They proposed the use of a sequence index and density ratio to measure the significance of a sequential pattern. An index $> 1$ implies a \textit{follow} relationship between two events, an index $< 1$ implies a \textit{repel} relationship, and an index $= 1$ implies an independent relationship. Link prediction is a popular method for predicting future relationships. \textit{Missing link prediction} considers the mere structural properties of a network to estimate unobserved relationships. \textit{Temporal link prediction} considers both structural and temporal properties of a network~\cite{acar2009link,menon2011link}. The predictive influence of temporal properties in evolving networks was originally viewed with less importance~\cite{yang2015evaluating}. However, many researchers are taking the view that temporal properties are of equal importance, if not more, than the structural properties~\cite{yang2019advanced}. Time properties were considered key predictors in continuously evolving customer relationship management (CRM)~\cite{ali2018crm}. In this work, we focus on temporal link prediction. In this emerging research area, a temporal matrix factorization approach (TMF) was first introduced~\cite{yu2017temporally}. However, TMF only considered network structure from the previous timestep. This work was extended by~\cite{yu2017link}, introducing LIST, which is both backward and forward-looking link prediction. LIST supports multiple timesteps, in the sense that the network state (snapshot) is observed at each timestep. The network, rather than the nodes within the network, progresses over time. LIST works on the assumption of node similarity. As a result, it does not work well with networks with evolving features over multiple timesteps. An example of such a network is business workflows with correlated but heterogeneous activities. 

We now review two related but distinguishable AI research areas. These are task recommendation and sequence prediction.

Reactive task recommendation is widely studied in autonomous agents modeling. Autonomous agents can make a deterministic or stochastic choice based on fixed or continuously changing contexts~\cite{albrecht2018autonomous}. In multi-agent systems, the context can include the acting agent, its opponents, or collaborators. Task recommendation models predict the next action based on a current state or past actions. A recommendation algorithm, SPEED, can predict the next activity of a smart home inhabitant based on past episodes of activities~\cite{alam2012speed}. Task recommendation can be formally defined as:
 $P(x_{(T+1)} \vert x_1 ,x_2,\dots,x_T)$. One of the challenges with the task recommendation is the modeling of changing behavior and context~\cite{albrecht2018autonomous}. Another limitation of the task recommendation is its lack of temporal consideration. A model may recommend turning the lights off after observing recent activities such as turning off TV and thermostat. However, the time distance of prior activities may increase the error rate of task recommendation models. Sequence prediction and dynamic network link prediction both incorporate timeliness in their predictive modeling. 
 
 \begin{table}[!b]
  \begin{tabular}{p{0.1\linewidth}p{0.15\linewidth}p{0.6\linewidth}}
    \toprule
    Node &Adjacency&Activity\\
    \midrule
    1 &	2,3	& Start\\
    2 &	3,4,7,9 &	Client initiated contact through walk-in, iLead, phone, Email, SMS, etc\\
    3 &	4,7,9 &	Actual contact though in-person, phone, Email reply or SMS reply\\
    4 &	5 &	Appointment set\\
    5 &	6 &	Appointment confirmed\\
    6 &	7,11 &	Appointment complete\\
    7 &	8,9,10 &	In-person visit\\
    8 &	9,10,12	& Test drive\\
    9 &	8,10,12 &	Deal negotiation\\
    10 &	4,11 &	Turn-over\\
    11 &	8,9,10,12 &	Be-back (Subsequent in-person Visit)\\
    12 &   & Deal Closed\\
  \bottomrule
\end{tabular}
\caption{CRM activities represented as nodes  }
\label{tab:activities}
\end{table}

A sequence prediction, formally defined as $P(y_1,y_2,\dots,y_{\hat{T}}| x_1,x_2,\dots,x_T)$\cite{sutskever2014sequence}, exploits hidden and observable states of a set of sequence data and maps it to the next sequence. Applications of sequence prediction include automated language translations, speech recognition, and user action sequence learning.

\section{Our Approach}\label{sec:methods}

A challenge in learning dynamic networks lies in the complexity of adapting to continuously evolving time and context~\cite{albrecht2018autonomous,yu2017temporally,menon2011link}. The influence of time is intuitively understood for a temporal network. However, the degree of influence depends on the network's sensitivity to granular temporal relations. Before we describe our approach, we start by describing the type of network that is mainly under consideration in this work (Subsection \ref{ssec:consistentnetworks}). We then discuss our a new approach of canonically representing heterogeneous time-evolving activities as a \textit{TP-matrix} (Subsection \ref{ssec:tp-matrix}). The limitation of the adjacency matrix is explained before we show the use of \textit{TP-matrix} is more suitable for evolving networks. In Subsection \ref{ssec:pi-matrix}, we demonstrate the retention of temporal relationships from the global network can be further enhanced using a time-parameterized predictive influence (\textit{TPPI}).  

\begin{figure}
  \center
  \includegraphics[scale=0.4]{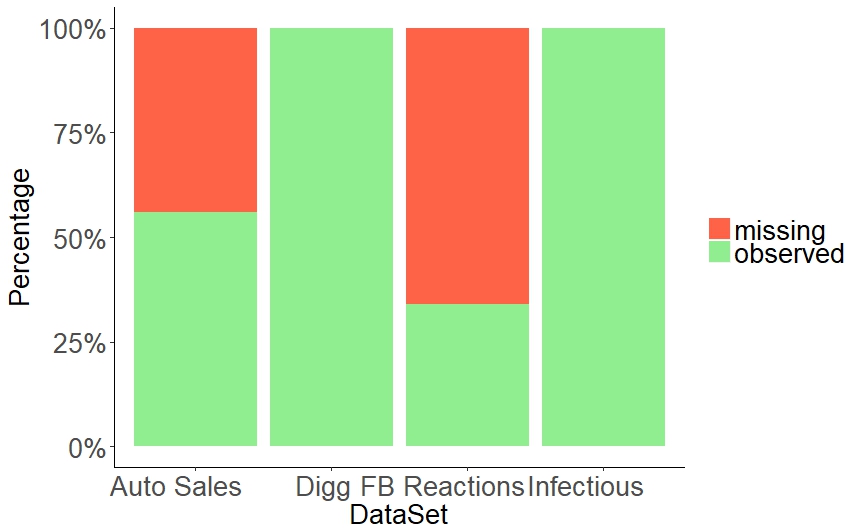}
  \caption{Missing vs observed edges for different networks}
  \label{fig:missinglinks}
\small
\end{figure}

\subsection{Temporally Consistent Networks}\label{ssec:consistentnetworks} 
Earlier in the paper, we mentioned that link prediction estimates the next relationship in a dynamic network. Temporally consistent networks are a special kind of dynamic networks that smoothly evolve or devolve based on an observed or hidden relationship~\cite{yu2017link}. Temporal consistency has its roots in physics, but it was recently used to improve predictions in image-based graphics and video sequencing~\cite{eom2019temporally}. We now formally define the properties that hold true for temporally consistent networks. \theoremstyle{definition}
\begin{definition}{\textbf{Temporally Consistent Networks}}
Let a sequence of network snapshots $\mathcal{G}$ represent the evolution of a temporally consistent network. The nodes in $\mathcal{G}$ 1) are temporally interconnected, 2) influence the network evolution, and 3) contribute to an internal or external outcome.
\end{definition} Examples of temporally consistent networks include activity networks, road traffic networks, and social networks. The evolution of phenomena can be predicted using the spatial, temporal, or thematic relationships of a temporally consistent network.

\subsection{Time-parameterized Matrix}\label{ssec:tp-matrix}
Time-parameterized matrix (\textit{TP-matrix}) is our resolution for class imbalance which hampers the effectiveness of link prediction models using an adjacency matrix~\cite{menon2011link}. The use of adjacency matrix or binary edge weights (0, 1) to indicate a link between two nodes limits the ability to model all states in a global network~\cite{yu2017temporally}. This limitation is particularly evident in forming networks with a large number of unobserved links. In addition, Figure \ref{fig:missinglinks} shows graphs of real-world events tend to be sparse. The absence of a link conveys more noise than information when binary edge weights are used~\cite{yu2017temporally}. We now explore the suitability of (\textit{TP-matrix}) for modeling heterogeneous, time-evolving networks.

\begin{figure*}[t]
  \center
  \includegraphics{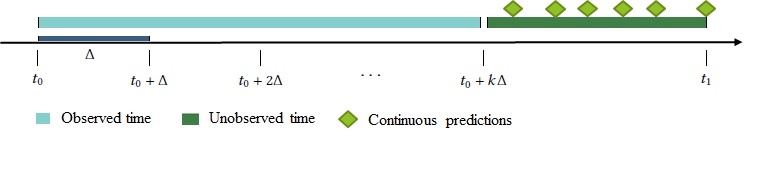}
  
  \caption{Dynamic network timeline with continuous predictions }
  \label{fig:networktimeline}
\small
\end{figure*}

We will illustrate our approach by using concrete examples from a real-world dataset of sales activities. We transform a sequence of observed activities and their timestamps into a network in the form of a \textit{TP-matrix}. Each row represents one instance of a sale. Each column represents an event or an activity during the lifetime of a corresponding sale. The values of the matrix represent timed edges $a_{(i,j)}$. In lieu of using binary edge weights, we use temporal residuals. Each value contains the transition timestamp from node $i$, the initial event of the $i^{th}$ instance of a sale to node $j$, which represents the corresponding $j^{th}$ activity. In the absence of a transition between node $i$ to node $j$, the value $a_{(i,j)}$ is the time distance between the initial event $i$ (i.e. the receipt of a sales request) and the running time $\Delta$. We use the term 'time-parameterization' with respect to the node-level timestamps and time distances. At the network level, the term 'temporal parameterization' is more appropriate as it emphasizes the relationship aspect of all temporal properties. We make this distinction to note that our reference to time-parameterization in this paper emphasizes the timing aspect whereas temporal parameterization emphasizes the evolutionary aspect of temporal relationship at the global network. The computation of the \textit{TP-matrix} is presented below. First, we use the running time $\Delta$ in Equation \ref{eqn:temporalmatrix_step1} to eliminate class imbalance due to missing edges.
\begin{equation}\label{eqn:temporalmatrix_step1}
 \left\{a_{(i,j)}\right\}_{j=1}^n ~where~ a_{(i,j)}=\left\{ \begin{array}{rcl}
|v_i-v_j |,& \mbox{if} & (v_i ,v_j  ) \in E \\
|v_i-\Delta |,& \mbox{if} & (v_i ,v_j  ) \not\in E
\end{array}\right.
\end{equation}
Second, we normalize the temporal values obtained in Equation \ref{eqn:temporalmatrix_step1} to $\mathbb{R} \in (0, 1]$. The normalization method is shown in Equation \ref{eqn:temporalmatrix_step2}. 

\begin{equation}\label{eqn:temporalmatrix_step2}
 \left\{a_{(i,j)} \right\}_{j=1}^n ~where~ a_{(i,j)}=\frac{1}{1+\abs{a_{(i,i)}-a_{(i,j)} }}
\end{equation}

This normalization maintains the non-zero scale to avoid re-introducing noise. It is similar to normalizing user-item ratings in collaborative filtering for recommendation systems~\cite{mu2018survey},  where each user's rating is divided by the maximum rating allowed. In our case, we cannot use the global denominator because temporal distances between activities are continuously variable. 

\subsection{Time-evolving Predictive Influence}\label{ssec:pi-matrix}
In subsection \ref{ssec:tp-matrix}, we looked at how we better capture the evolution of network structure through time. In this section, we will concentrate on the evolution of time and its predictive influence. The distinction between structure-function and time-function relationships are often overlooked, but it is a significant research problem for dynamic networks. Figure \ref{fig:networktimeline} shows a dynamic network timeline split into observed and unobserved parts. The current state of the art in link prediction considers the evolution as a function of continuous snapshots. A fixed time interval between timesteps must be defined during training. This means the prediction can happen only at the same fixed interval. A model trained with a 2-day fixed interval expects the same 2-day interval at prediction time. This is not ideal for practical applications because a random, on-demand prediction is not possible. An alternative approach would be to make the time interval user-configurable rather than fixed. However, that would exponentially increase the number of trained models. To make a continuous prediction possible as illustrated by the green diamonds in Figure \ref{fig:networktimeline}, it is important to separate or decouple the time-function relationship from the structure-function relationship. We develop a time-parameterized predictive influence (\textit{TPPI}) to capture the influence of time in a temporal relationship.

We start by evaluating the contribution of a given node to the evolutionary progress of its network.

\theoremstyle{definition}
\begin{definition}{\textbf{Time-Parameterized Predictive Influence}}
Given a node $v_i \in V_t$ at time $t$ and a predictive influence threshold $\beta \in [0,1)$, the predictive influence of $v_i$ is:
\begin{equation}\label{eqn:predictiveinfluence}
 TPPI(v_i,\beta)=P( v_i | f(v_i)(1-\beta)
\end{equation}
\end{definition}

The predictive influence of each node is maximized given the node's feature vector $f(.)$ using Equation \ref{eqn:influencemaximization}.

\begin{equation}\label{eqn:influencemaximization}
P( v_i | f(v_i)) = \underset{f}{\mathrm{max}} \prod_{j=i-\alpha}^{i+\alpha} P( v_j | f(v_i))
\end{equation}
Backward and forward-looking maximization is used over the temporal space in between $T-\alpha$ and $T+\alpha$, where $\alpha$ is a tunable hyperparameter, which allows for the adjustment of the temporal degree of the neighborhood during model training. \textit{TPPI} learns from feature vectors of each pair of nodes ($v_i, v_j) \in V_t$ and their temporal weights $a(i,j)$ obtained from Equation \ref{eqn:temporalmatrix_step2}, yielding the following Equation \ref{eqn:predictivesigmoid}.

\begin{equation}\label{eqn:predictivesigmoid}
P( v_j | f(v_i))= \frac{1}{1+e^{-f(v_i)f(v_j)a(v_i,v_j)}}
\end{equation}

Note that a higher influence threshold $\beta$ makes \textit{TPPI} more conservative. A value of $0$ is equivalent to using no threshold. The threshold hyperparameter helps balance or fine-tune the temporal and structural context sensitivities of the network being modeled.   
\begin{figure*}[t]
  \center
  \includegraphics[scale=0.7]{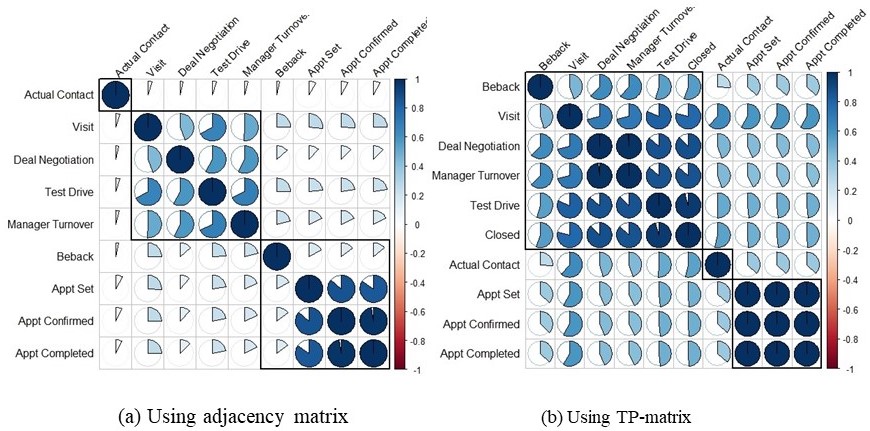}
  
  \caption{A graphical display of a correlation matrix among sales activities using (a) adjacency matrix and (b) TP-matrix. All activities show positive correlations to one another because their coefficients are greater than 0 and shaded blue with more color intensity and a filling pie as they get closer to 1. }
  \label{fig:correlationmatrix}
\small
\end{figure*}  

\subsection{Comparative analysis of variance and correlation}
The correlation matrix  using Pearson correlation coefficients is effective for showing pair-wise correlations of more than two independent variables. The matrix values are correlation confidence values in the interval [-1, 1]. A smaller or larger absolute value, respectively, indicates a weaker or stronger pair-wise correlation, whereas the negative or positive (-/+) sign shows the direction of the correlation. For predictive influence and feature selection analysis, the maximum confidence value of $1$ indicates a collinearity problem where the pair-wise relation of two variables is so high they are likely identical. A confidence value of $0$ shows no linear relationship.

Now, we will show \textit{TP-matrix} can capture the spatiotemporal correlation among network nodes relatively better than the adjacency matrix. The graphical representation of the correlation matrix for sales activities are shown in Figure \ref{fig:correlationmatrix}. The Pearson correlation with hierarchical clustering is shown for both the adjacency matrix and \textit{TP-matrix}. We can observe several benefits of the TP-matrix over the adjacency matrix. The adjacency matrix shows a weak intra-node correlation because of the sparsity of the graph data. The \textit{TP-matrix} conveys more spatiotemporal correlation with more realistic clustering. For example, a vehicle test drive, a deal negotiation, a referral to a sales manager (manager turn-over), and a conclusion of a deal (closed) usually all happen in one initial or subsequent visit to a dealership showroom. The clustering of these activities in (b) of Figure \ref{fig:correlationmatrix} is, therefore, more realistic. In addition, we can observe that it is not possible to calculate the pair-wise correlation coefficient between $closed$ and any other activity using the adjacency matrix. Since the dataset included successful deals only, an adjacency matrix would have the value $closed = 1$ for all instances. As a result, a variance $\sigma^2$ and, therefore, a Pearson's correlation coefficient cannot be calculated. This supports our claim (and that of others~\cite{yu2017temporally}) that the adjacency matrix cannot model all states in a global network. We summarize our findings below.

\begin{enumerate}
\item Using the running timestep $\Delta$ eliminated a key limitation due to missing edges when using the adjacency matrix. With this approach, the absence of a link conveys latent feature variance and relevance with respect to observed links. 
\item Using timed edges improved the retention of temporal relationships with respect to the global network. It also highlighted more granular relationships among sub-groups of nodes. The rectangles in (b) of Figure \ref{fig:correlationmatrix} show clusters of correlated activities that matched the mental model of automotive experts. For example, independent activities that deal with appointment scheduling and tracking are in the same cluster. During deal negotiation, a sales manager may get involved. Such activity is tracked as 'Manager Turnover'.
\item With this approach, the model variance is reduced because 1) the model is less prone to the cold start problem at the initial phase when the network is forming, and 2) the model regulates itself with a decay function as the network grows larger. We will expand on this in the next subsection. 
\end{enumerate}

\begin{table}[!b]
\begin{tabular}{|p{0.5cm}|p{0.7cm}||p{0.5cm}|p{0.5cm}|p{0.7cm}||p{0.5cm}|p{0.5cm}|p{0.7cm}| }
\hline
\multicolumn{2}{|c|}{$D(t)=e^{-\theta(T-t)}$} &
\multicolumn{6}{|c|}{D(t) = Eq. (\ref{eqn:decayfunction})} \\
\multicolumn{2}{|c|}{Example 1} &
\multicolumn{3}{|c|}{Example 1} &
\multicolumn{3}{|c|}{Example 2} \\
\hline
\hline
  Node & $t_2$-$t_1$ & $t_1$ & $t_2$ & $t_2$-$t_1$ & $t_1$ & $t_2$ & $t_2$-$t_1$\\
\hline
$v_1$ & - & 0.5 & 0.4 & $\downarrow$ & 0.5 & 0.4 & $\downarrow$ \\
 \hline
$v_2$ & - & 0.8 & 0.7 & $\downarrow$ & 0.8 & 0.7 & $\downarrow$ \\
 \hline
$v_3$ & - & 0.1 & 0.0 & $\downarrow$ & 0.1 & 0.4 & $\uparrow$ \\
 \hline
$v_4$ & - & 0.7 & 0.6 & $\downarrow$ & 0.7 & 0.7 & $\rightarrow$ \\
\specialrule{1pt}{1pt}{1pt}
 $\theta$ & 6 & 0.525 & 0.425 & $\downarrow$ & 0.525 & 0.55 & $\uparrow$ \\
 D(t) & 0.548 & 0.622 & 0.563 &  & 0.622 & 0.638 &  \\
\specialrule{1pt}{1pt}{1pt}

\end{tabular}
\caption{A few example results of different decay functions }
\label{tab:decayexamples}
\end{table}

\subsection{Regularization Method}
Regularization is a technique used to reduce model variance. In its simplest form, a uniform decay $\frac{1}{2}$ can be applied to an objective function. A much popular approach is to use a decay function with a parameter $\frac{1}{2}\theta$. The parameter regulates how fast the decay hastens. In temporal link prediction, an exponential decay function $D(t)=e^{-\theta(T-t)}$ with time $t$ $\in T$ is commonly used to regulate the importance of the current network snapshot compared to previous snapshots~\cite{yu2017temporally,yu2017link,yang2019advanced }. 

As the network progresses, the farthest network states become less relevant than the current network state. The acceleration of the irrelevance (decay) depends on the problem and can be estimated by the decay parameter $\theta$. However, one can argue that an exponential decay function for a temporal link prediction is not necessarily a strictly monotonically decreasing function. This is especially true for temporally consistent networks. Let us consider the citation network, which is sensitive to temporal evolution. On the question of the most impactful papers, aging papers with continuous and active citations may have higher relevance than recent papers without proven staying power. This supports our argument that previous network states are not necessarily less relevant than the current state. We can take this argument further by pointing out that, on a more granular level, certain nodes may continue to contribute to a network's evolution at a much higher rate than other nodes. With this in mind, we define a relative exponential decay function in (\ref{eqn:decayfunction}) by first taking the Mean \textit{TPPI} (See Equation \ref{eqn:predictiveinfluence}) of the current network state. A lower Mean \textit{TPPI} infers  more nodes that are not contributing to the network evolution. We then use it as the decay parameter since it better reflects a node-level temporal decay of the network state rather than a mere time decay.     

\begin{equation}\label{eqn:decayfunction}
D(t) = e^{-(1-\theta)} ~where~ \theta = \frac{1}{d}\sum_{i=1}^d  TPPI(v_i,\beta)
\end{equation}

Table \ref{tab:decayexamples} shows synthetic examples of a network decay with four nodes. We assume that $T-t = 0.1$ and $\theta = 6$ for the commonly used decay function. With our relative exponential decay, you can see that an over-performing node (in terms of its predictive influence) can slow down the temporal decay of the overall network, even when the influence of other nodes are diminishing. In contrast, the commonly used decay function is a function of time decay and, therefore, a strictly decreasing function overtime.   

\begin{table}[!b]
\begin{tabular}[c]{|p{1cm}||p{3.6cm}|p{2.8cm}|}
\hline
 Parameter Name & Description & Average Degree\\
 \hline
 $\alpha$ & The temporal range of neighborhood (e.g. 2, 3 or more +/- hops) during training and inference. & \textbf{Optional}\newline Default value: 3\\
 \hline
 $\lambda$ & The learning rate or step size used during optimization. & \textbf{Optional}\newline Default value: 0.1\newline
 Valid values: A positive float value from 0.1 to $10^{-4}$.\\
 \hline
 $\beta$ & Predictive influence threshold. A higher value gives more weight to the structural context, whereas a lower value gives more weight to temporal context. & \textbf{Required}\newline Valid values: positive float value in (0, 1]\\
 \hline
 $\gamma$ & The momentum for the stochastic gradient descent (SGD) algorithm. & \textbf{Optional}\newline Default value: 0.9\newline
 Valid values: A positive float value in (0, 1]\\
 \hline
 $M$ & Maximum number of iterations. & \textbf{Optional}\newline Default value: $10^3$\newline
 Valid values: A positive integer value\\
 \hline
\end{tabular}
\caption{Hyperparameters for the TPNM model}
\label{tab:hyperparameters}
\end{table}
\subsection{TPNM Model Training}
We trained our model using stochastic gradient descent (SGD) with momentum~\cite{ruder2016overview}. The momentum $\gamma$, learning rate $\lambda$, and the temporal range $\alpha$ are tunable, user-defined hyperparameters that are initially set (as default values) to 0.9, 0.1 and 3, respectively. We chose these default values because they are  widely accepted as initial values and also worked well for our model. The complete list of hyperparameters is in table \ref{tab:hyperparameters}.

The model is iteratively trained until it converges. Convergence is achieved when the minimum loss variation is consistently below a tolerance threshold of $10^{-3}$ for the last 10 epochs. More specifically, the training stops when the stop condition in Equation \ref{eqn:stop-condition} is satisfied, where $E$  is the error vector produced in a maximal number of boosting iterations $M$.

\begin{equation}\label{eqn:stop-condition}
\abs{\sum_{i=M-13}^{M-10} E_i-\sum_{i=M-3}^{M} E_i}<10^{-3} 
\end{equation}
Given a \textit{TP-Matrix} $\mathbf{A}$(t) and \textit{TPPI-matrix} $\mathbf{R}$(t) at time $t$, we minimize the following objective function for $\mathbf{U}$ and $\mathbf{V}$.

\begin{equation}\label{eqn:objective}
\begin{aligned}
J(U,V) = \frac{D(t)}{2}\sum_{t=1}^T (\mathbf{A}(t)-\mathbf{R}(t)^\top )^2\\
+\frac{1}{2}\lambda \sum \norm{\mathbf{U}}_F^2+\frac{1}{2}\lambda \sum \norm{\mathbf{V}}_F^2
\end{aligned}
\end{equation}

We carried out most of the time-dependent optimization legwork in subsection \ref{ssec:pi-matrix}. The first term is regulated with the relative exponential decay function D(t), whereas the other two terms are regulated by a learning rate $\lambda$ that is continuously reduced from its initial value to $10^{-4}$.

We show the derivatives of the objective function below.

\begin{equation}\label{eqn:errorU}
\frac{\partial{J}(U, V)}{\partial{U}} = 
\sum_{t=max(1, T-\alpha)}^{T}d\sigma(\mathbf{U}.\mathbf{V}^\top) - \sum_{i=0}^d\frac{\lambda}{2} \mathbf{U}
\end{equation}

\begin{equation}\label{eqn:errorV}
\frac{\partial{J}(U, V)}{\partial{V}} = 
\sum_{t=max(1, T-\alpha)}^{T}d\sigma(\mathbf{U}.\mathbf{V}^\top) - \sum_{i=0}^d\frac{\lambda}{2} \mathbf{V}
\end{equation}

The pseudocode for our model is presented in Algorithm \ref{alg:algorithmTPNM}.

\begin{algorithm}[h]
\caption{Temporally Parameterized Network Model}
\label{alg:algorithmTPNM}
\SetAlgoLined
\textbf{Input}: $G = (V, E)$ in the form of a \textit{TP-matrix} as defined in Equation \ref{eqn:temporalmatrix_step2}\;
\KwResult{Learned parameter matrices U, V  }
 Initialize matrices $U, V, U_{last}, V_{last}$ with random values\;
 \While{stop condition (\ref{eqn:stop-condition}) is not satisfied}{
  Compute $\frac{\partial{J}(U, V)}{\partial{U}}$ and $\frac{\partial{J}(U, V)}{\partial{V}}$ using Equations \ref{eqn:errorU} and \ref{eqn:errorV}\;  
  \If{$\lambda > 10^{-4}$}{
   Reduce $\lambda$ by 0.05\;
   }
  Update $U = U - \frac{\partial{J}(U, W)}{\partial{U}} + \gamma U_{last}$\;
  Update $V = V - \frac{\partial{J}(V, W)}{\partial{V}} + \gamma V_{last}$\;
  $U_{last} = U$\;
  $V_{last} = V$\;
  Update error vector $E$ with $J(U,V)$ using Equation \ref{eqn:objective}\;
 }
 \algorithmicreturn{ U, V} \;
\end{algorithm}

\section{Experiment Setting}\label{sec:experiments}
While this work involves link prediction, we are interested in different types of graphs that have not been addressed in most link prediction studies. In the absence of related studies we could use as a baseline, we first start our experiment with networks that are traditionally used for link prediction problems. We evaluate how our approach works on two datasets used in temporal link prediction and then expand our evaluation to more complex networks.

 \begin{table}[t]
\begin{tabular}{|p{1.5cm}||p{1.6cm}|p{1.4cm}|p{2.6cm}| }
 \hline
 
 Dataset & Total Nodes & Average Degree & Absent:Observed Edge Ratio\\
 \hline
Infectious & 410 & 84 & 0\\
 \hline
Digg & 30,398 & 21.5 & 0\\
 \hline
 CRM \break Activities   & 120,000    & 3.5 &   11:14\\
 \hline
 Facebook Reactions & 2,088 & 4   & 691:353 \\
 \hline

\end{tabular}
\caption{Properties of the experimental dataset  }
\label{tab:graphstats}
\end{table}

\subsection{Dataset}
We used a total of four datasets of dynamic networks. Two of them, Infectious and Digg, were part of KONECT\footnote{http://konect.uni-koblenz.de/networks/} networks, which are widely used in link prediction. These two datasets were used for baseline comparative analysis. The third dataset was a collection of social reactions on Facebook posts from the British Broadcasting Corporation (FBReactions\footnote{https://github.com/naman/fb-posts-dataset}). It contained 348 Facebook posts on a wide range of topics with at least one of 6 possible social reactions (\textit{like, love, wow, haha, sad, angry}). Missing edges (e.g., missing reaction type \textit{angry}) were high (66\% of all edges). The published social reactions dataset did not include timestamps on social reactions. A timestamp was randomly generated for each reported social reaction. Posts without at least one social reaction were removed from consideration. Since Facebook uses the same 6 reactions across their social media platform, this experiment should apply to Facebook Live videos as well. The fourth was a real industry dataset (AutoSales) collected from a live automotive CRM~\cite{ali2018crm}. This dataset contained 10,000 anonymized sales events, each representing a network with 12 nodes. Missing edges (e.g., missing activity type \textit{sales appointment}) accounted for 44\% of all possible edges. Table \ref{tab:graphstats} shows the properties of the datasets.

\begin{table*}[h]
\centering
\begin{tabular}{p{1.6 cm}||p{2.6 cm}p{2.6 cm}p{2.6 cm}p{2.6 cm}p{2.6 cm}p{2.6 cm}}
\toprule
    {Method} & {Infectious} & {Digg} & {FBReactions} & {AutoSales} \\
    \hline
\midrule
    WCN & 0.9719$\pm$0.8300 & 0.2719$\pm$0.2064 & Not evaluated & Not evaluated \\
    HPLP & 0.5883$\pm$0.6767 & 0.2702$\pm$0.2034 & Not evaluated & Not evaluated \\
    CPTM & 0.8847$\pm$0.9367 & 0.2196$\pm$0.2469 & Not evaluated & Not evaluated\\
    TMF&0.5309$\pm$0.1185 & 0.0032$\pm$0.0011 & 0.5475$\pm$0.2205 & 0.5107$\pm$0.9288 \\ 
    LIST & 0.3824$\pm$0.1114 & \textbf{0.0026$\pm$0.0005} & 0.5206$\pm$0.0008 & 0.4759$\pm$0.4803 \\ 
    TPNM (ours) & \textbf{0.1681$\pm$0.6295} & 0.1129$\pm$0.6420 & \textbf{0.1338$\pm$0.94610} & \textbf{0.0783$\pm$0.1611} \\
\bottomrule
\end{tabular}
\caption{Average RMSE of TPNM in comparison to baseline methods}
\label{tab:results}
\end{table*}

\subsection{Experiment Design and Evaluation Metric}
Like most link prediction algorithms, our work is a regression task. It uses the structure and time relationships within a network to learn its evolution. Root-Mean-Square Error (RMSE) is used to evaluate the generalization capability of regression-based algorithms. In this paper, we will report both the iterative (epoc steps) and overall RMSE performance. Mean absolute error (MAE) is also reported as an additional performance evaluation. In addition, we show the training run-time performance with six different network samples. The graph representation allows for evaluating the accuracy performance of regression-based link prediction using the area under the curve (AUC) binary measurement. AUC has been used extensively as an evaluation metric for link prediction problems~\cite{yu2017temporally}. However, this requires the preparation of the test data in a certain way. We now describe our approach using the automotive CRM dataset.

CRM activities were converted into a sequence of states and their corresponding timestamps $(x_0,t_0),\dots,(x_i,t_i), ~where~ \forall i \ t_i < t_{i+1}$. For uniformity during training and evaluation, each state or activity was assigned an identifier, as shown in Table \ref{tab:activities}. The temporal property of each activity was then captured by a \textit{TP-matrix}. The goal of this experiment setup was to enable the evaluation of our link prediction approach against our baseline, TMF, using AUC. For example, 1200 different networks, with ten different nodes each, were tested. For any network, we ran link predictions at most $T-1$ stages. That is for every node in the network at any given time except the destination node at T. Our training data was the activity sequence data represented as a network at timestep $T$. For our test data, we removed the node at $T$. We considered the test positive if the predicted node matched with the node at $T$. Otherwise, we considered it negative. We can now use the binary measurement and compare our results with our baseline TMF.

\begin{figure}[t]
  \center
  \includegraphics[scale=0.35]{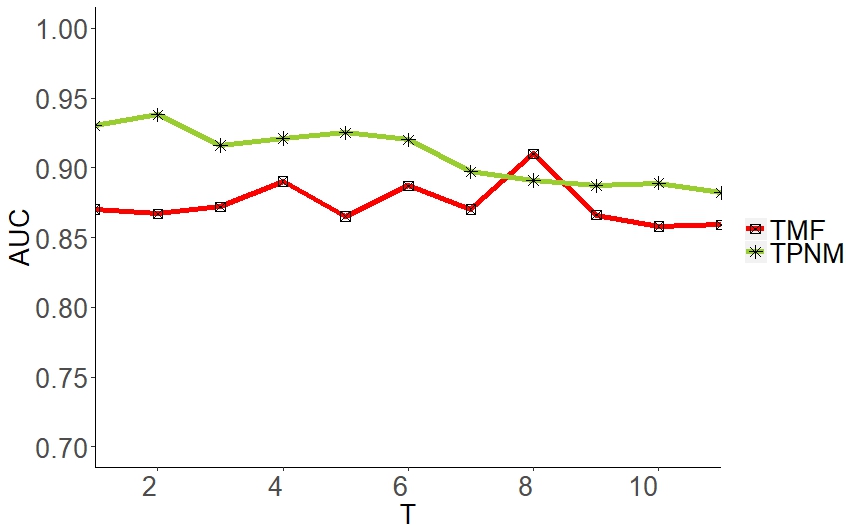}
  \caption{AUC performance over running time compared to TMF (baseline)}
  \label{fig:aucperformance}
\small
\end{figure}

\section{Results}

We compared our method with the recent link prediction benchmark results using four datasets. Table \ref{tab:results} summarizes the RMSE performance of TPNM compared to the baseline method (TMF) and four additional methods. Interestingly, the Digg dataset did not perform well using our approach. This could be explained by the weak temporal relations in user ratings. Digg is a dataset that contains stories and their ratings (how many users dig a given story) at a given time. It can be argued the temporal span of ratings does not affect the outcome (e.g., whether a story is going to trend or not). More specifically, the timing of ratings and how they are scattered over the lifespan of a story may not have more predictive influence than the ratings. Therefore, our approach may not work for networks that are less sensitive to temporal evolution than they are on topological evolution. On the contrary, we observed that our model performed well for the activity or social reaction networks because timing is more critical for these networks.

Using AUC, TPNM obtained a better overall accuracy than the baseline TMF as shown in Figure \ref{fig:aucperformance} with 93\% compared to 87\%. In this case, we used the automotive CRM dataset only. The fluctuation of the baseline method is reduced. We believe the temporal retention across nodes (discussed in Section \ref{sec:methods} improved the accuracy of the link prediction.

\begin{figure}[t]
  \center
  \includegraphics[scale=0.38]{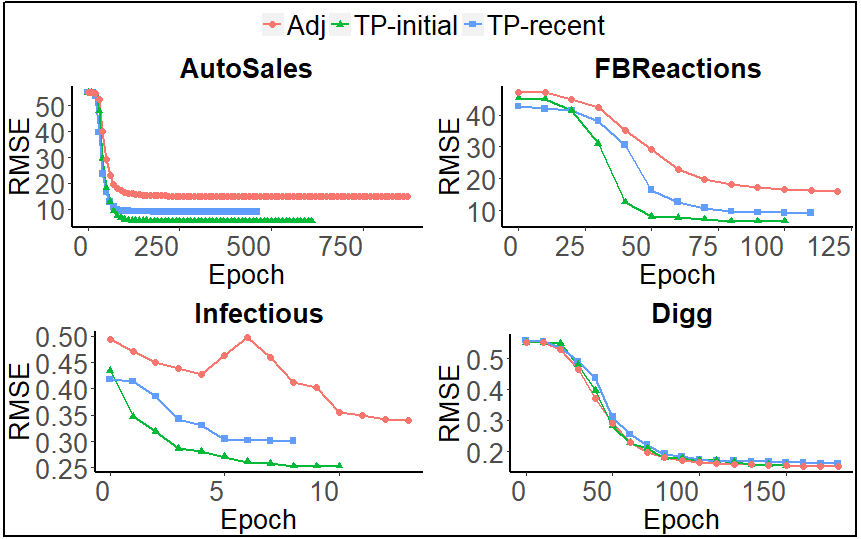}
  \caption{RMSE performance while training with four datasets using adjacency matrix (red), TP-matrix using the most recent event (blue) and, TP-matrix using initial event (green)}
  \label{fig:rmseperformance}
\small
\end{figure}

We now extend our evaluation to the RMSE performance results obtained from an iterative model training. Three experiments were carried out to evaluate the training performance of TPNM using \textit{TP-matrix}. In the first experiment (Adj), we simply used the adjacency matrix. In the second and third experiments, we used \textit{TP-matrix} with the initial (i.e., the network conception) event of the network influencing all other events (TP-initial) or with the most recent event influencing the current event (TP-recent). An example of computing \textit{TP-matrix} using the initial event was explained in Equation \ref{eqn:temporalmatrix_step1}. Using the most recent event means simply rewriting $v_i-v_j$ to $v_{j-1}-v_j$ in the same equation. Our findings, as shown in Figure \ref{fig:rmseperformance}, showed that the use of \textit{TP-matrix} improved the RMSE performance compared to the adjacency matrix. The use of \textit{TP-matrix} using the initial event (TP-initial) performed better than when the most recent event (TP-recent) is used. That is because the initial event provides a more stable temporal consistency among all nodes (i.e., a smooth network evolution) over the most recent event. The curve stops when our iterative training obtains the least RMSE rate, which was 0.07 for the activity dataset.

\begin{figure}[t]
  \center
  \includegraphics[scale=0.4]{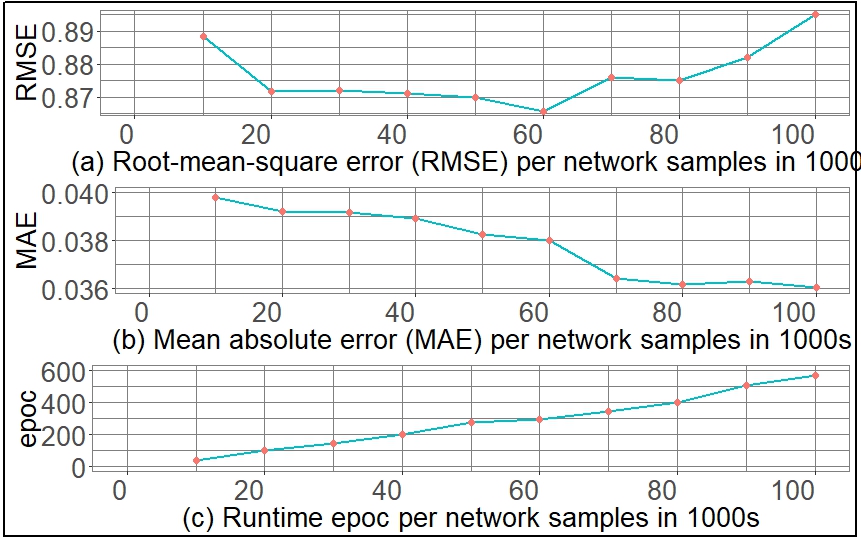}
  \caption{Training performance per number of networks sampled.}
  \label{fig:performancepersamples}
\small
\end{figure}

Finally, Figure \ref{fig:performancepersamples} shows the overall iterative training performance using RMSE and MAE. It also shows the runtime of the TPNM algorithm as a function of the size of the network instances. For 100,000 networks with over 1.1 million nodes, the training completed just under 10 minutes.

\section{Applications}\label{sec:applications}
Many real-world tasks involve time-sensitive steps whose timely or untimely completion affects the desired outcome. Time-parameterization of link prediction opens the doors to link prediction applications that were otherwise hard to apply. In the following lines, we discuss a few promising applications of our approach. 

\paragraph{Business Workflow Optimization} Optimizing a workflow using intelligent decision points requires context-sensitive learning based on what activities occurred at what time, what activities could happen at the current state, and how that could affect the desired outcome of a workflow. This presents the challenges of heterogeneity and dynamicity that we discussed earlier. In sales workflows, activities are correlated but heterogeneous, and paths to a sale (i.e., the desired outcome) are dynamic and depend on the network state. More specifically, in automotive sales workflow, a deal negotiation may be recommended early for online consumers, because they are more likely to complete their product research on the Web before they even initiate contact with a seller. The same activity may be recommended much later for in-store consumers. Therefore, timing and other side information (e.g., deal negotiation) may present boosting or diminishing predictive influence. Our new approach of representing heterogeneous activities as a temporally parameterized network and retaining temporal relationships with respect to the global network enables researchers to extend link prediction applications to business workflow optimization.

\paragraph{Viral Marketing} Link prediction was recently used for influence maximization in viral marketing~\cite{zhu2018location}. For product promotion, influence users in social networks are grouped based on their in-degree (followers) and out-degree (friends) relationship. The idea is to predict the next social connection of a target group for marketing campaigns. Another viral marketing trend includes Facebook's live video advertisement. Social reactions to a Facebook live video are essentially time-evolving activities. Our new approach can be used to predict the optimal time to show an advertisement in a live video feed with active social reactions.

\paragraph{Talent Acquisition Recommendation} Talent acquisition, unlike recruitment, is an evolving strategy to actively and passively attract top talent. Similar to the business workflow, timing is critical when identifying top candidates that are showing signs of job market activities. Job market events and candidate activities can be learned as a network to predict the optimal time to engage a prospective talent.

\section{Conclusions}\label{sec:conclusions}
In this paper, we presented a new approach, TPNM, to link prediction that enables and motivates the interdisciplinary application of artificial intelligence and dynamic networks for new domain areas, including business workflow optimization, viral marketing, and context-aware action recommendation. We reviewed the limitation of using the adjacency matrix in link prediction and presented a new time-parameterized matrix as a solution for class imbalance and feature noise due to missing edges. Our research showed that networks that model real-world events or activities contain a high proportion of missing links. In addition, we showed that these events or activities tend to be heterogeneous with dissimilar features. Our empirical evaluation showed that the use of a time-parameterized matrix with a predictive influence index is more suitable for heterogeneous time-evolving networks. We used two published datasets to compare our approach to existing link prediction methods. In addition, we tested our approach with two additional datasets for heterogeneous sales activities and social reactions. For all the datasets we evaluated, we achieved an RMSE smaller than 0.17. For time-evolving datasets, the results were much better with 0.13 and 0.07 RMSE for social reactions and automotive sales activities respectively.

\section*{Acknowledgment}
This research was partially funded by the University of Missouri-Kansas City, School of Graduate Studies.

\vfill\null 


\bibliographystyle{IEEEtran}




\end{document}